# An AIC-based approach for articulating unpredictable problems in open complex environments

Haider Al-Shareefy
ECS
University of Southampton
Southampton, UK
hal1e20@soton.ac.uk,
haideralshareefy@outlook.com

Michael Butler
ECS
University of Southampton
Southampton, UK
m.j.butler@soton.ac.uk

Thai Son Hoang
ECS
University of Southampton
Southampton, UK
t.s.hoang@soton.ac.uk

*Abstract*— **This research paper presents an approach to enhancing the predictive capability of architects in the design and assurance of systems, focusing on systems operating in dynamic and unpredictable environments. By adopting a systems approach, we aim to improve architects' predictive capabilities in designing dependable systems (for example, ML-based systems). An aerospace case study is used to illustrate the approach. Multiple factors (challenges) influencing aircraft detection are identified, demonstrating the effectiveness of our approach in a complex operational setting. Our approach primarily aimed to enhance the architect's predictive capability.**

## I. INTRODUCTION

To provide a higher level of assurance of the trustworthiness of systems designed to operate in open, complex environments, an architect needs a deeper understanding of challenges in such complexity. Typically, Machine Learning (ML)-based systems are deployed in such environments. For example, an interesting incident occurred in Los Angeles whereby a food delivery robot was reported to wheel through a crime scene [1]. The incident video shows a standby cameraman lifting the yellow tape and allowing the robot to drive through the crime scene. In this system's interaction, an emergent unintended relationship has formed between the robot and the cameraman. The robot and the bystander's primary purposes have somehow cooperated, prompting the human to adopt an emergent unexpected auxiliary purpose, whereby the news reporters appreciated the robot's action by permitting the robot's behaviour to wheel through the crime scene. Our research investigates a different way to solve the uncertainty problem for system design by repurposing systems approach to enhance architect predictive capability.

For our case study, we have chosen an aircraft collision avoidance problem from the work done in [2] in which the team briefly analysed the problem and made an engineering judgment about the type of factors to include to describe the operational environment of the perception system, weather, aircraft type, Time of Day Geographic region. After analysing the problem comprehensively using our approach, we discovered several factors influencing the aircraft detection problem, some of which were related to the type of factors we need to consider for operational problem data.

Our contribution, therefore, is a systematised series of predictive thinking operations based on the Appreciation, Influence, and Control (AIC) framework. This systematisation transforms abstract concepts from systems theory into actionable steps for predictive analysis. These steps allow for a structured approach to unravel complex, dynamic interactions in systems, making it easier for the architect to predict and mitigate potential issues or conflicts.

## II. BACKGROUND

General System Theories (GST) provide a comprehensive framework for understanding and analysing complex systems. These theories conceptualise systems as interrelated and interconnected entities that function as part of a larger whole. [3]. They offer principles and models for understanding how systems behave, interact, and influence each other. This understanding is crucial for predicting how a system might respond to changes in environmental or internal dynamics. Chain-of-Thought CoT techniques in Artificial Intelligence and Natural Language Processing involve explicating the reasoning process step-by-step, akin to how humans approach problem-solving. This method allows for breaking down complex tasks or scenarios into smaller, more manageable steps [4] [5]. By applying CoT to systems theory, architects can sequentially dissect a complex system, examine its components, and understand the interdependencies and interactions. This approach makes the reasoning process more transparent and interpretable.

An example of a system theory is AIC [6] which we used to establish a systems approach for systems engineering practice. AIC-based General Systems Approach (GSApp) [7] can be viewed as a systems thinking blueprint or general pattern for developing systems engineering processes to solve complex problems. The GSApp is based on adapting the Appreciation, Influence, Control (AIC) GST [8]. This paper uses adapted AIC concepts as a fundamental systems perspective for understanding and predicting systems. It defines a system as a structure of interconnected components with a purpose, actions, and capabilities. The approach described in [7] categorizes systems into agents, non-agents, and environmental types, each driven by a Primary Purpose (PrimeP), which is assumed to be unchanged across different contexts.

III. ENHANCING ARCHITECT PREDICTIVE CAPABILITY USING AIC-BASED CHAIN-OF-THOUGHT TECHNIQUE

The study in [5] delves into how prompting techniques enhance the ability of large language models to perform complex reasoning tasks. In our research, we realised that when we use CoT techniques with General System Theories, we can methodically work through the complexities of open operational environments. This methodical approach allowed us to predict and justify our assertion about potential outcomes based on interactions and changes between the system and its operational environment. Every step in our CoT method consists of a predictive question that triggers the thinking operation and a predictive guiding prompt that assists with solving the question. Building on this foundation, our research introduces the adoption of the AIC framework as an integral component of the Chain-of-Thought (CoT) process. The AIC-based Chain-of-Thought (AIC-based CoT) process applies the AIC systems approach to improve confidence in predictive reasoning. Therefore, the solution of every step is an assertion claim: "The architect asserts that …". The steps in the process are iterative, which means the architect seeks clarification and help to answer the question and solve the prompt while refining their prior assumptions made in the earlier step. Architects may carry on in each step until no new information can be added. We will present the process by defining the general steps to be taken and explain how each step was applied in the case study. A full version of the analysis can be found in [9]

*1) Identify a list of unsafe behaviours:*

**Consider the following predictive question:** Given the input information provided in the problem brief, what is the list of all possible unsafe appreciative, influence or control interactions? **Guiding prompt:** Describe the unsafe behaviours in the observed systems phenomenon. The step is complete. when the architect judges that all the unsafe behaviours have been identified.

**Application to collision avoidance case study:** The guiding prompt helped us identify a control-based unsafe behaviour. In this case, the unsafe behaviour is the unplanned convergence of flight paths between two or more aircraft. The next steps will help us to predict all appreciative, influence, and control factors related to the unsafe behaviour. **The output of this step is the architect prediction:** The architect asserts that the initial observation of the problem is the following unsafe behaviour: two or more aircraft come into unplanned contact during flight. One aircraft is equipped with a computer perception-based collision avoidance system. The intruder aircraft is following a flight path that intersects with another aircraft.

*2) Observe and identify the systems contributing to the unsafe behaviour:*

**Consider the following predictive question:** Given the output from step 1, what are the systems involved in the unsafe behaviour? **Guiding prompt:** Considering the unsafe behaviour as a system of systems, identify the systems that contribute to the unsafe behaviour. The step is considered complete when all contributing systems have been identified. **Therefore, the architect's prediction for the collision avoidance case study would be:** The architect asserts that the following is a valid list of purposed systems of concern: perception-based collision avoidance system (AVP), the intruder aircraft, and the ownship aircraft.

**Revisiting earlier steps:** Sometimes, analysis in a step leads to the need to revise earlier steps. For example, while considering the impact of the intruder aircraft motion on the perception system, we realised that an extra assumption needs to be added to the output of step 1, namely that the intruder aircraft is in constant motion. Thus, we revise the output of step 1 to the following: *The intruder aircraft is in constant motion, following a flight path that intersects with another aircraft.*

*3) Define actions of the contributing systems that cause unsafe behaviour:*

**Consider the following predictive question:** Given the output from steps 1-2, what are the unsafe appreciative, influencing, or control actions? Guiding prompt: Infer the identified systems' immediate unsafe actions that contribute to the unsafe interaction. The unsafe action provides an answer to what makes the overall interaction unsafe. The step is considered complete when the architect judges that all system actions have been defined.

**Application to collision avoidance case study:** Taking the computerized perception-based mid-air collision avoidance system (AVP), as an example, the guiding prompt directed us to infer the AVP's immediate action that is contributing towards the unsafe interaction. The output of step 1 does not explicitly define what the AVP is exactly doing. Thus, we inferred that the perception model failed to detect the intruder aircraft as a most likely contributing action to the overall unsafe interaction. **Therefore, the architect's prediction for the collision avoidance case study would be:** The architect asserts that the only potential unsafe action of AVP is the failure of the perception model to detect an intruding aircraft.

*4) Determine the primary purpose behind the unsafe action:*

**Consider the following predictive question:** Given steps 1-3 outputs, what is the contributing system's original primary purpose (PrimeP)? **Guiding prompt:** Envision that the identified systems have a purpose to master a skill that mitigates the unsafe action. With that in mind, define systems' primary purpose (PrimeP) as if it has the intent to govern or master such a skill. A purpose can be defined as a verb that describes a system-level action done by a system unto another system. One implicit assumption (which must be made explicit in the architect's assertion statement) is that a system's PrimeP is fixed in any given scenario. This assumption is crucial, as it forms the basis for justifying and predicting AIC auxiliary purposes in any other context. In other words, systems' auxiliary interactions may change based on the situation at hand. However, the driving PrimeP remain constant regardless of the situation. Therefore, it is important that the architect chooses a PrimeP carefully. The step is considered complete when the architect judges that all PrimePs have been derived.

**Application to collision avoidance case study:** Considering the context of the overall unsafe interaction and the definition of the unsafe contributing action of AVP failing to detect intruder aircraft, we inferred that, ultimately, the AVP's primary purpose would be to acquire the capability of detecting an intruder aircraft. **Therefore, the architect's prediction for the collision avoidance case study would be:** The architect asserts that the one and only primary purpose of the AVP system is to govern the capability to detect intruder aircraft reliably and correctly. There is no other possible primary purpose besides that in any given scenario.

*5)   Predict auxiliary Influence interaction:*

**Consider the following predictive question:** Given the output from steps 1-4, which other system, capability, or behaviour must the identified systems indirectly control to achieve their respective PrimePs? **Guiding prompt:** For each system, determine an auxiliary indirect influence purpose to achieve the respective PrimeP. Once the purpose of auxiliary influence has been identified, a list of influential actions should be determined. Recognizing the subtle difference between a direct control action and an indirect influence action is important. Influence action is an action from the source system unto an aspect that is outside the source's sphere of possible direct control and within the target-influenced sink's sphere of possible direct control. On the other hand, a control action would be an action that the source system performs on an aspect within its own sphere of direct control to achieve an influence action. Therefore, when choosing an influence action, it is important to consider something about the sink that can only be indirectly controlled by the source system. The step is considered complete when the architect judges that all auxiliary influence interactions have been identified.

**Application to collision avoidance case study:** The guiding prompt provided us with a characterisation to identify what other aspect is outside of the AVP sphere of possible direct control, related to the intruder aircraft and important to influence to achieve the primary purpose.

**Revising step 4**: This CoT has led us to question the practicality and appropriateness of the chosen AVP's PrimeP being "to detect intruder aircraft". The latter PrimeP seemed to be too thought-restraining for a general purpose, which prompted us to reconsider the AVP's primary purpose from a different perspective, which would make the initial PrimeP an auxiliary step towards it. With this in mind, we redefined the AVP's primary purpose to be *"assisting the ownship pilot in appreciating the open airspace safety"*. From such a perspective, we can identify the following as an auxiliary influence purpose *"enhancing ownship pilot's decision-making process"* to be a reasonable assumption.

When defining an influence action, consider acting upon AIC aspects that affect the sink's behaviour. One aspect that controls the pilot's decision-making process is their "situational safety awareness". Therefore, the action that can achieve influence is *to augment the pilot's situational safety awareness of their surroundings*. The step is considered complete when the architect judges that all influence actions have been identified.

**Given the above analysis, the architect's prediction for the collision avoidance case study:** The architect asserts that a valid auxiliary influence purpose for AVP to achieve its primary purpose, the (AVP) must at least aim to enhance the (own_aircraft_pilot_decision_making _process). To achieve the purpose of influence, AVP must at least perform the following influence action: AVP augments the pilot's awareness (own_aircraft_pilot_situation_ awareness) of their surrounding environment (surrounding_ airspace_safety).

*6)   Predict auxiliary Control interaction:*

**Consider the following predictive question:** Given the output from steps 1-5, which capability, system, or behaviour should the identified systems aim to control to achieve their respective auxiliary influence purpose? **Guiding prompt:** Consider every influence action as the auxiliary control purpose, then define a list of control actions that deliver the control purpose. The step is considered complete when the architect judges that all auxiliary control interactions have been identified. The step is considered complete when the architect judges that all possible control behaviours have been identified.

**Application to collision avoidance case study:** Step 6 defines specific control actions and capabilities the AVP system should possess and employ to fulfil its auxiliary influence action of augmenting the pilot's awareness of their surroundings. Direct control involves controlling aspects within the AVP's sphere of direct control. By employing a collision threat predictive model to anticipate future positions of potential intruder aircraft and evaluate collision risks, the AVP system can generate effective avoidance strategy recommendations, thereby exerting indirect control over the pilot's situational awareness.

**The architect's prediction for the collision avoidance case study would be:** The architect asserts that a valid **control purpose** would be to enhance (own_aircraft_pilot_decision_ making_process). To achieve the valid control purpose, AVP must at least perform a control action of employing (threat_predictive_model) to forecast subsequent positions of (intruder_aircraft_position) and evaluate the risk of a potential collision (risk_of_potential_collision).

*7)   Predict auxiliary Appreciation interaction:*

**Consider the following predictive question:** Given the output from steps 1-6, what other systems must the identified systems appreciate to ensure the success of their control behaviours in delivering the required control purposes? **Guiding prompt:** For every control action, infer the appreciation purpose of some third-party appreciated system, which impacts the success of the control action in manifesting its control purpose. Appreciated system behaviours directly impact the identified system control behaviour. The step is considered complete when the architect identifies all possible appreciated systems and appreciative actions.

**Application to collision avoidance case study:** This step involves understanding the limitations and external factors that the system must appreciate. In the case of the AVP system, an appreciative interaction would be related to the correct and reliable detection of intruder aircraft in a variety of attitudes. Also, appreciating the impact of environmental factors like sunlight on its visual sensors is crucial for ensuring accurate detection and tracking of potential threats, thereby supporting its primary and auxiliary control purposes of predictive threat modelling. Therefore, **the architect's prediction for the collision avoidance case study:** The architect asserts that in order to ensure the effectiveness of the identified control behaviour in achieving the intended impact of the control purpose, the AVP must acknowledge the context and potential limitations of visual information received from imaging sensors (own_aircraft_camera). The AVP needs to acquire the following appreciative behaviour: account for (sun_position) in the sky relative to the direction of the camera (own_camera).

*8) Predict and analyse factors and challenges:*

**Consider the following predictive question:** Given the output from steps 1-7, what are the factors or challenges involved in the problem domain, the most influential factors or challenges, and potential sources of surprising emergence? **Guiding prompt:** highlight all possible factors or challenges (systems and capabilities) involved in the situation from the predicted knowledge. After collating all factors or challenges, define each factor and compute its frequency of mentioning in the analyses. The most mentioned factors or challenges are the most influential factors or challenges. However, the least mentioned are not the last to worry about, they indicate potential red flags for sources of potential surprising emergence. The step is considered complete when all factors are captured, defined and evaluated for frequency. **Application to collision avoidance case study:** We collated all the factors and challenges mentioned throughout the steps 1-7 analyses by capturing the factors in (x_y_z) format. We then computed each factor or challenge's frequency to understand their influence and significance. The most frequently mentioned factors are considered the most influential in the problem domain, while the least mentioned ones may indicate potential red flags or areas for surprising emergence. **Architect prediction:** The architect asserts that there are 104 influential factors involved in the problem. Furthermore, the architect asserts that Own_aircraft_pilot_decision_making_process (11 mentions) is the most critical factor in the problem domain.

## IV. REFLECTION

The paper explores AIC-based problem articulation for system design. We used Appreciation, Influence, and Control sense-making based General Systems Approach as the basis to formulate problem articulation CoT to solve a prediction task. Upon comparing our approach to Smyers [2] team's method of problem articulation to identify influential factors on the perception system, we discovered additional factors to consider by using AIC-based CoT: the position of the sun relative to the camera, drones, birds, intruder aircraft attitude (pitch, roll, yaw), intruder aircraft distance, ownship aircraft (pitch, roll, yaw), military aircraft intrusion, cloud turbulence, fog, humidity and raindrops shapes and sizes. Indicating that the approach improved our predictive capabilities as we foresaw more potential factors that affect the perception performance, for example, a possible hack into the system that controls the AVP detection threshold or delays the warning alert, thus more problem scenarios to add to the test exhaustiveness. As well as other factors that affect the avoidance system functionality as a whole (not just its perception).

Other factors were discovered that were not directly related to affecting the false positives or negative rates of the mid-air collision problem as a whole. For example, potential hack on detection threshold, wind effect on camera, and accelerated dynamics of ownership aircraft. Such factors should have been mentioned in the initial problem brief.

## V. RELATED WORK

The closest framework to our approach we found was Critical Systems Heuristics (CSH) [10], an approach that focuses on critically evaluating and understanding complex situations or systems. Also, work done by Shannon [11] Metathinking is a general critical and reflective approach to thinking that involves making predictions through 4 different metathinking modes about the problem: structural, process, relational, and transformational. Barclay [12] discusses the importance of problem articulation using systems thinking for systems design as he recognizes the system problem as a wicked problem.

## VI. FUTURE WORK

Our future work will further examine the architect's theoretical side and how it will be involved in V&V arguments for system trustworthiness. We envisage a paradigm shift focusing on enhancing the Architect's predictive capability, analogous to enhancing a super-intelligent language model to handle system uncertainty.

**Acknowledgements**: Al-Shareefy is supported by a Thales EPSRC iCASE Award. Butler and Hoang are supported by the HD-Sec project, funded by the Digital Security by Design (DSbD) Programme delivered by UKRI to support the DSbD ecosystem.